\documentclass[aps,prd,onecolumn,12pt,preprintnumbers]{revtex4-1}

\usepackage{hyperref}

\usepackage{graphicx}
\usepackage{epstopdf}

\begin{document}

\preprint{HDP: 16 -- 05}





\title{Air modes of the Bacon internal resonator banjo}

\author{David Politzer}

\affiliation{California Institute of Technology,
452-48 Caltech,
Pasadena CA 91125}

\date{August 8, 2016}

\begin{abstract}
Sound measurements on a sequence of related, similar constructions with slightly different dimensions confirm a simple picture of the air modes of the internal resonator banjo's body.  For the purpose of this study, the air modes are decoupled from the soundboard (i.e., [drum] head) modes by replacing the head with $3/4''$ plywood.  The resulting characteristic features survive the strong coupling of the air modes to the head and are in accord with the qualitative distinctions recognized by banjo players.

\bigskip


\bigskip

\bigskip

**************************







\bigskip

\noindent {\bf contact info:} politzer@theory.caltech.edu, (626) 395-4252, FAX: (626) 568-8473

\bigskip


\bigskip

\bigskip

\bigskip

\bigskip

\bigskip

\bigskip

\bigskip

\bigskip

\bigskip

\bigskip

\bigskip

\bigskip

\centerline{1}

\end{abstract}

\maketitle{}

\section{Introduction}

\medskip
\noindent 

Virtually every banjo design ever made continues to have enthusiasts and remains in production to this day.  Almost all fall into one of two categories: open-back or resonator.  Of the alternatives and variations introduced over the past century and a half that are neither, the ``internal resonator," first patented\cite{patent} and put in production by stage performer Fred Bacon in 1916, is something of a cult favorite.  The originals are coveted, and the basic design is still produced by independent luthiers.

Banjos are stringed instruments whose body is a drum and whose string-head interface is a floating bridge.  The most common drum is a cylinder (called the ``rim"), with the head stretched taut over its top edge.  If that top edge is made out of a separate piece, that edge piece is called the ``tone ring."  Tone rings can be hardwood, but more often they are metal.  They can be solid, hollow, or a combination.  They range from simple ${1 \over 2}$ lb rings made of ${1\over4}''$ diameter brass rod to elaborate cast and machined combinations, weighing up to $3 {1\over2}$ lbs.  Open-back banjos are just that: the drum has no back or bottom.  However, the player's body effectively forms a back, and the sound hole is the space between rim and body, where air can enter from and escape to the ambient surroundings.  Resonator banjos have a wooden back.  However, it does not seal against the bottom edge of the rim.  Rather, there is a small space all around which serves as a sound hole.\cite{resonators} 

The novel feature of Bacon's design is the internal structure of the drum.  The purpose of this study is to gain a qualitative understanding of the acoustics of that structure.  Like all banjos, almost all the sound is produced by the vibration of the head.  Subtle variations arise from the interaction of the head with the internal air pressure variations.  The coupling is strong because the head is light and flexible and makes contact with the internal air over its whole area.  Bacon's and subsequent internal resonator banjos used the standard head designs of their times.  And the head/internal air coupling is not particularly amenable to detailed study or analysis. The focus here is on the internal air itself.

The drum (or ``pot") of the internal resonator banjo has a partial back attached to the rim and an additional internal wall that divides the pot interior into a central cylinder that has an open back and an enclosed annular region, as diagramed in FIG.~1.  

\begin{figure}[h!]
\includegraphics[width=3.2in]{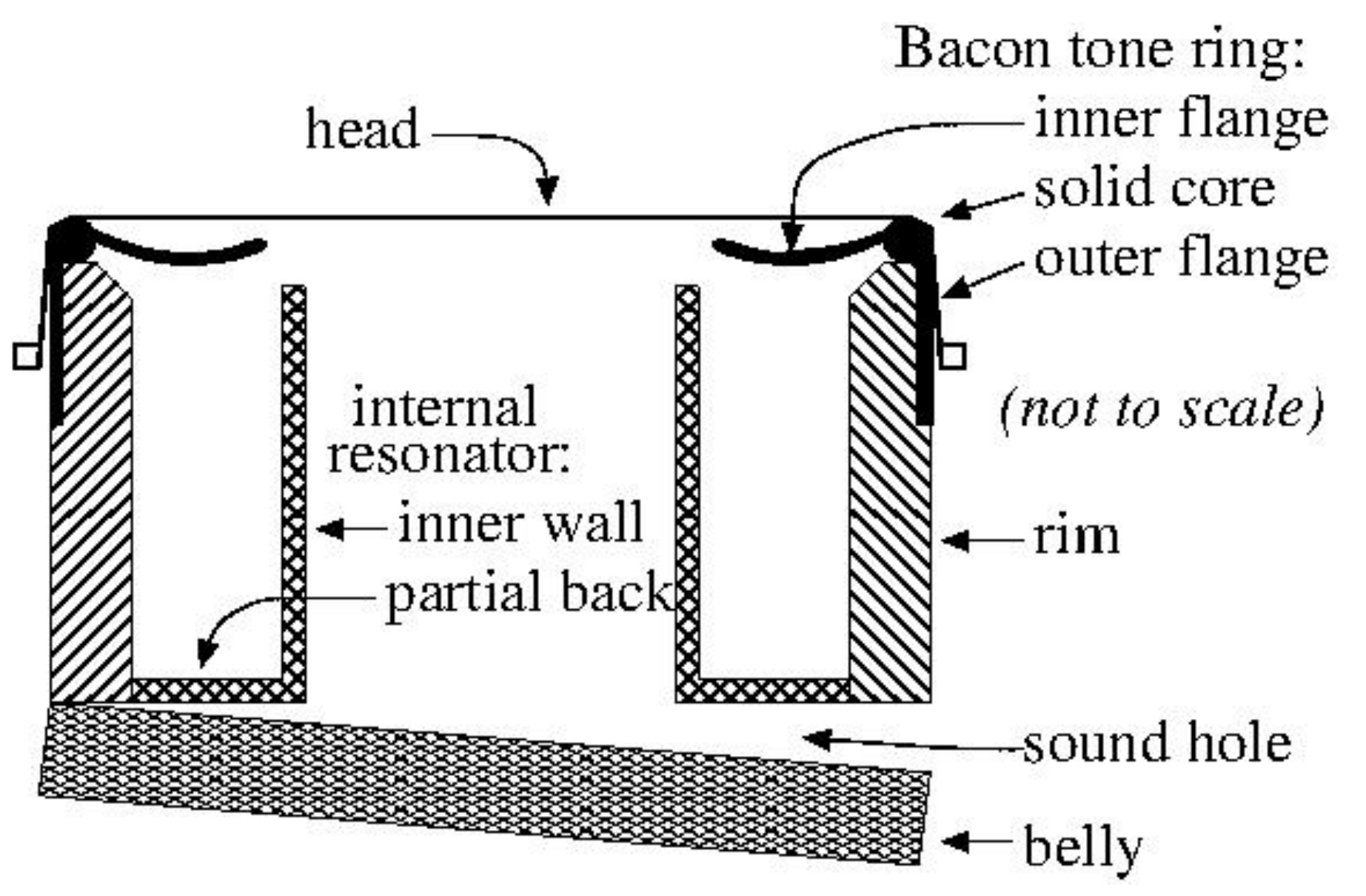}
\caption{Schematic cross section of the internal resonator pot}
\end{figure}

\begin{figure}[h!]
\includegraphics[width=3.5in]{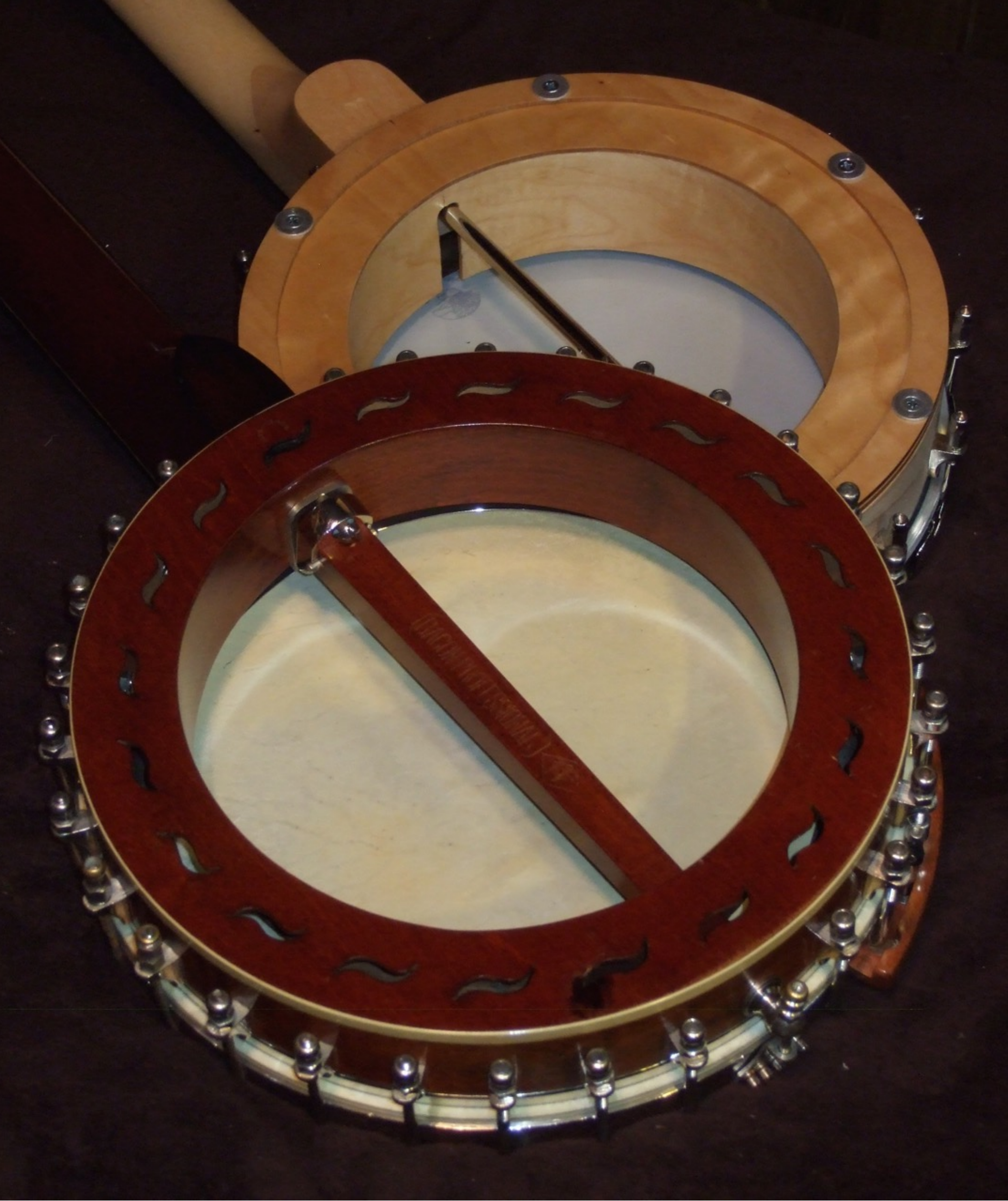}
\caption{Back view of an original Bacon Professional ff and the modified Deering Goodtime}
\end{figure}
\noindent There is a small space between the top edge of the inner wall and the head that serves as an air connection between the two regions.  Bacon's original models also had metal tone rings, weighing about 1 lb, of a design that has come to be known by his name.  Those tone rings have a solid (or almost solid) core, with a sheet metal wrapping that forms horizontal and vertical flanges.   The vertical, outer flange is fitted snug to the outer side surface of the wood rim, while the inner, horizontal flange is free to vibrate.  That inner flange also further defines the air passage between the central cylinder and the outer annulus.  When played, the central cylinder is partially closed by the player's body, and the sound hole is the whole passage between the partially enclosed interior cylinder and the outside air.

Even Fred Bacon himself was not particular about the exact dimensions or even design details.  In the early years of his company, he sold banjos made in part or entirety by leading manufacturers of the time.  Many of those instruments exist to this day, and they vary considerably in their details.  However, they all fit the description given above.  What was studied for the analysis that follows is a new banjo with interchangeable internal resonators of different dimensions and identical banjos with and without a Bacon-style tone ring.  FIG.~2 is a photo of an original ``Professional ff" (the Bacon Banjo Company's name for this design) and a modified Deering Goodtime banjo, equipped with an internal resonator and replica Bacon tone ring.  On both, the edge of the metal tone ring inner, horizontal flange is just barely visible.

\section{Summary of Results}

As with most stringed instruments, the lowest body resonance is a Helmholtz resonance, whose frequency is determined by the enclosed volume and the geometry of the sound hole.  The partial back provided by the internal resonator design decreases the interface area between the enclosed volume and the Helmholtz bottle ``neck."  The partial back also increases the volume of that ``neck."  Both of those features lower the resonant frequency.  (If it were a bottle of ideal geometry with neck-to-interior-volume interface area $A$ and neck volume $V_{\text{neck}}$, the frequency would be proportional to $A/\sqrt{V_{\text{neck}}}$.)  In spite of there being two internal volumes separated by a constriction, for geometries relevant to banjo construction, these volumes act as one --- at least with respect to Helmholtz resonance physics.  There is a single Helmholtz resonance whose frequency depends only on the size of the hole in the partial back and not on the height of the internal wall that separates the inner cylinder from the outer annulus.  Apparently, the area of the interface between the two regions is too large and the volume of the interface is too small for the system to behave anything like the coupled Helmholtz oscillators envisaged by Rayleigh as a logical possibility.\cite{rayleigh}  An ideal Helmholtz resonator has a frequency that is independent of the shape of the main volume.  So it is consistent that the one observed Helmholtz resonance frequency is independent of the internal wall height.

The closed-body air resonances are identified as coupled versions of the separate central cylinder and outer annulus.  The cylinder modes are calculable from the dimensions.  The annulus, while not exactly soluble, has modes that are well-approximated by periodic boundary conditions applied to a straight pipe with the annulus' cross section and effective circumference as its length.   Hence, the consequence of the wall is two-fold.  The lowest resonant  frequency of the annulus is significantly lower than that of a cylinder without the wall. (The longest annulus wavelength is essentially $\pi$ times the diameter, while  the longest wavelength of the no-wall pot is shorter by a factor of $\sim 1/1.84$,  coming from the zero of the appropriate Bessel function derivative).  And once the frequency spectra of the annulus and inner cylinder overlap, the combined system is has a greater density of resonances in frequency, which gives a more even response to driving frequencies.

For most banjos, the lowest half-octave or so is an example of a ``missing fundamental" in comparing the radiated sound to the plucked string frequencies.  The internal resonator  gives more support to these lower frequencies than present in the standard design.

The final element of Bacon's design is the tone ring itself.  In general, a metal tone ring provides a harder and stiffer edge to the vibrating area of the head than is presented by the same pure wood rim without it.  This means less dissipation.  Usually, this is most apparent at high frequencies (i.e., the ``hardness" aspect).  However, the net effect of the Bacon tone ring  is mostly the opposite because of the horizontal, vibrating flange.  The flange dissipates a noticeable amount of sonic energy that, in its absence, would have gone into radiated sound.  This is most apparent at the high ringing resonances of the flange itself.  Frequency-dependent damping means that the timbre of the transient sound of a pluck changes with time.  Discerning players refer to the ``finish" of a note, and some people particularly like the finish of the Bacon ring in which the highest frequency overtones die off more rapidly.  The flanges provide extra stiffness to the rim that reduces flexing in their respective planes.  That reduces energy loss at low frequencies.\cite{tone-ring}

In summary, Bacon's modifications offer a banjo with response stronger at low frequencies, smoother across all frequencies, and more subdued at high frequencies, especially as a sweeter finish to pluck sounds.  Those were his goals\cite{patent} and are consistent with the verbal characterizations offered by modern players.

\section{Internal Resonator Helmholtz Resonance}

Investigation of the Helmholtz resonance(s) begins with comparing partial backs with different size holes.  The bottom of a Bacon-tone-ring-equipped Goodtime rim was cut flat and fitted with six threaded inserts.  The rings shown in FIG.~3 were cut from 3.0 mm, 7-ply birch and could be attached with a narrow retaining ring of the same plywood and six screws.

\begin{figure}[h!]
\includegraphics[width=4.5in]{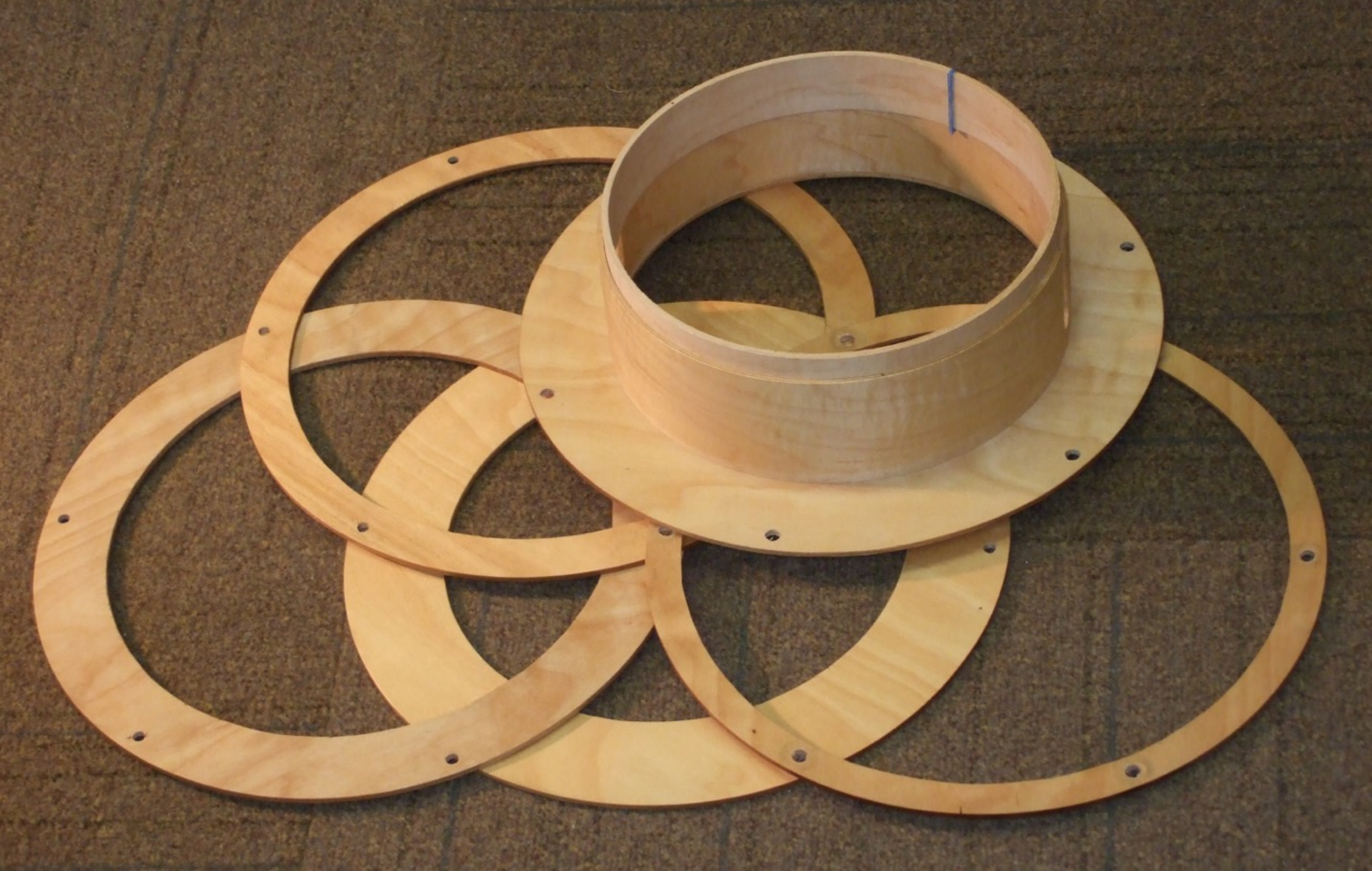}
\caption{Partial backs with various hole diameters and the adjustable-height internal resonator}
\end{figure}

With strings, neck, and tailpiece removed (but coordinator rod in place), the sounds of  head taps with a piano hammer were recorded for various bottom hole diameters.   The largest was the stock Goodtime, whose inner diameter is $9 {3\over4}''$.  The smallest was $7 {5\over8}''$, which is the diameter of the internal resonator insert  that appears later.

I mounted a synthetic belly, made of closed cell foam, cork, and Hawaiian shirt on the back  to serve as an approximation of the how the banjo is normally played.  Those materials were chosen to mimic the absorption and reflection of the player's body.    The opening to the outside air was chosen to approximate typical playing and is far more reproducible than holding the instrument up to one's body.  The genesis and details of this back are discussed in ref.~\cite{openback}.  Since a Helmholtz resonance is characterized by motion of air in and out of the sound hole, I placed the microphone right at the largest portion of the opening.  

FIG.~4 shows the spectra for long series of those head taps, plotted for 100 to 500 Hz.  

\begin{figure}[h!]
\includegraphics[width=6.5in]{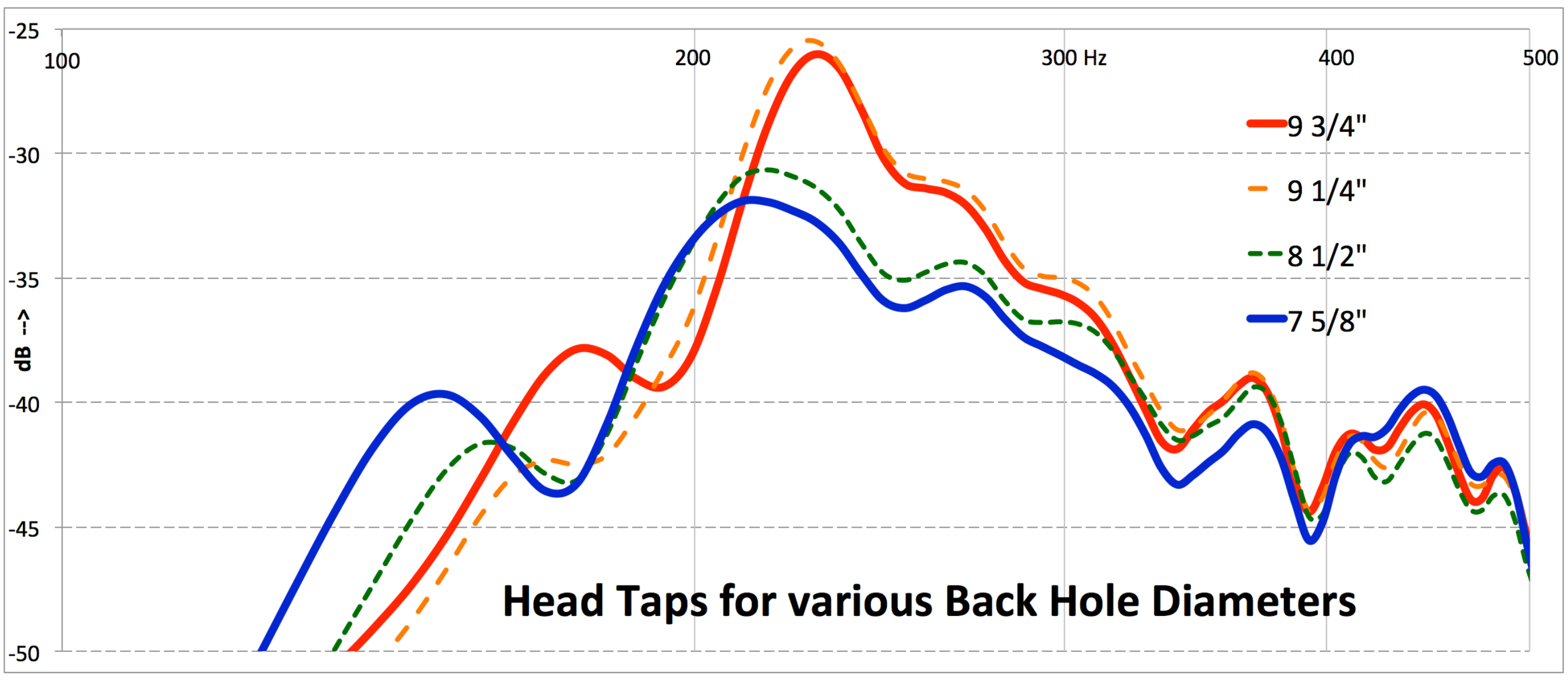}
\caption{Head taps with foam belly-back; curves labeled by back hole diameter (no internal wall)}
\end{figure}
The two lowest peaks show a systematic decrease in frequency with decreasing back hole diameter.  All higher frequency features show no appreciable frequency dependence on the back hole dimension.

This is the qualitative behavior expected from the Helmholtz resonator formula.  Smaller hole diameter implies smaller $A$ and larger $V_{\text{neck}}$.  There are two peaks for each back that reflect this behavior because the internal pot Helmholtz resonance couples strongly to the lowest drum mode of the head.  Not only do they both push the same plug of air in and out, but they also push on each other over the whole head surface.  The higher frequency modes are due to other physics.  It is typical that the lowest two modes of the body of a stringed instrument are the coupled versions of the Helmholtz and lowest sound board modes.\cite{rossing}  Note that on typical banjos the fundamental frequencies of all strings but the short $5^{\text{th}}$ string are below 300 Hz.

\begin{figure}[h!]
\includegraphics[width=4.5in]{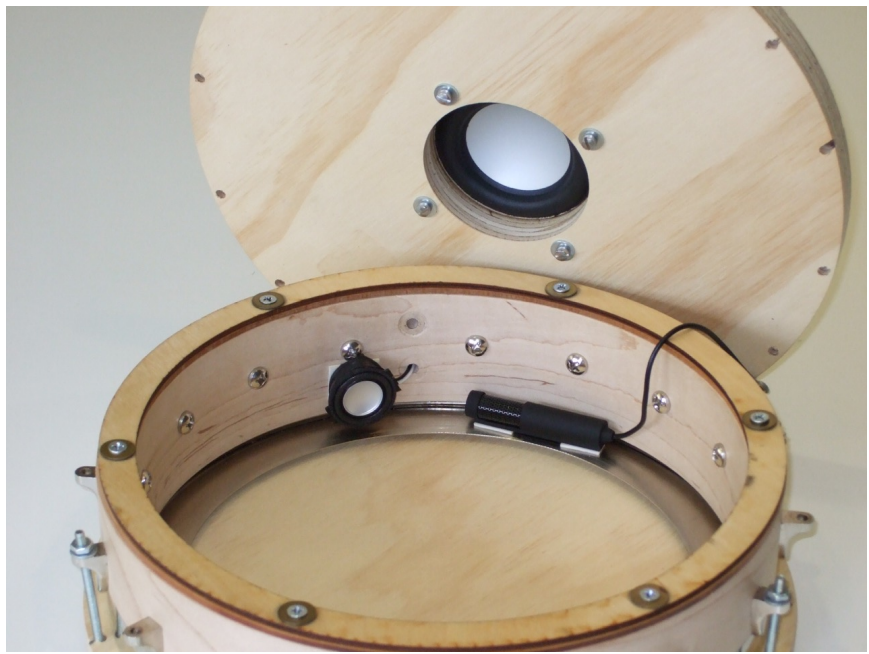}
\caption{Plywood heads, speakers, \& mic}
\end{figure}

To separate Helmholtz from sound board physics on violins and guitars, experimenters have occasionally buried the instrument in sand --- to immobilize the sound board motion.  It is easier on a banjo.  I replaced the regular mylar head with ${3\over4}''$ plywood.  To drive the Helmholtz resonance, I mounted a $3''$ speaker in the middle of that plywood head.  That head is the one {\it not} attached to the rim in FIG.~5 and installed on the rim in FIG.~6.  (The attached solid head and rim-mounted $1''$ speaker and microphone in FIG.~5 are described in section IV.)  The speaker is driven with a signal generator and audio amplifier with a slow sweep, logarithmic in frequency, over the desired ranges.

\begin{figure}[h!]
\includegraphics[width=3.5in]{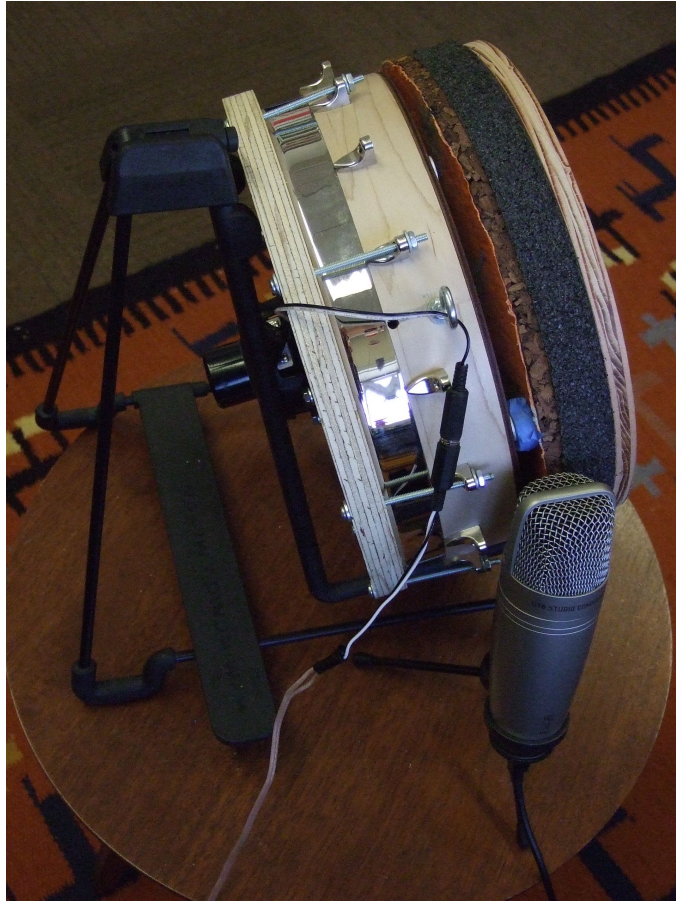}
\caption{Wood ``head" with $3''$ speaker and cork \& foam belly-back}
\end{figure}

So, the frequencies of the lowest two pot resonances are substantially lowered by the partial back in a way whose physics is qualitatively understood.  The next question is the impact of the cylindrical wall of the internal resonator that divides the interior into a smaller, central cylinder and an outer annular volume.  I fabricated a variety of internal resonators, all with the same cylinder diameter and back hole size but with various wall heights.  The cylinders were cut from 3.0 mm, 5-ply maple drum shell stock to produce wall heights ranging from ${1\over4}''$ to $2 {1\over4}''$.  However, the key to understanding the Helmholtz resonance and the cavity resonances (section IV) turned out to be internal walls of the roughly standard height, $2 {1\over4}''$, and {\it higher.}  To this end, I fabricated an adjustable height insert.   It had a split ring that could be inserted into a $2 {1\over4}''$  high cylinder.  The top edge of the split ring could be placed carefully at any particular distance from the head when assembled and tightened snugly with a shim in the gap in its circumference.  That is the upper right construction in FIG.~3.

The resulting spectra for driving with the $3''$ head-mounted speaker and listening with a microphone at the rim-belly-back opening (as shown in FIG.~6) are plotted in FIG.~7.  Now, for each pot geometry, there is only one, low, broad peak between 200 and 300 Hz.  With this set-up, the higher resonances are all considerably weaker.  All versions are with the same rim with its Bacon tone ring.  The curve labeled ``stock" refers to the standard, open-back Goodtime rim.  The curve labeled ``$7 {5\over8}''$" ring, is the partial back with no cylindrical wall.  (That $7 {5\over8}''$ is the same partial back hole size as the internal resonator.)  The dashed and dotted curves refer to internal resonators that have a ${3\over8}''$ and  ${1\over4}''$ space, respectively, between the top of the internal cylinder and the inner surface of the head.  The ``no gap" curve refers to an inner cylinder that touches the head and seals off the outer annulus from the inner cylinder.

\begin{figure}[h!]
\includegraphics[width=6.0in]{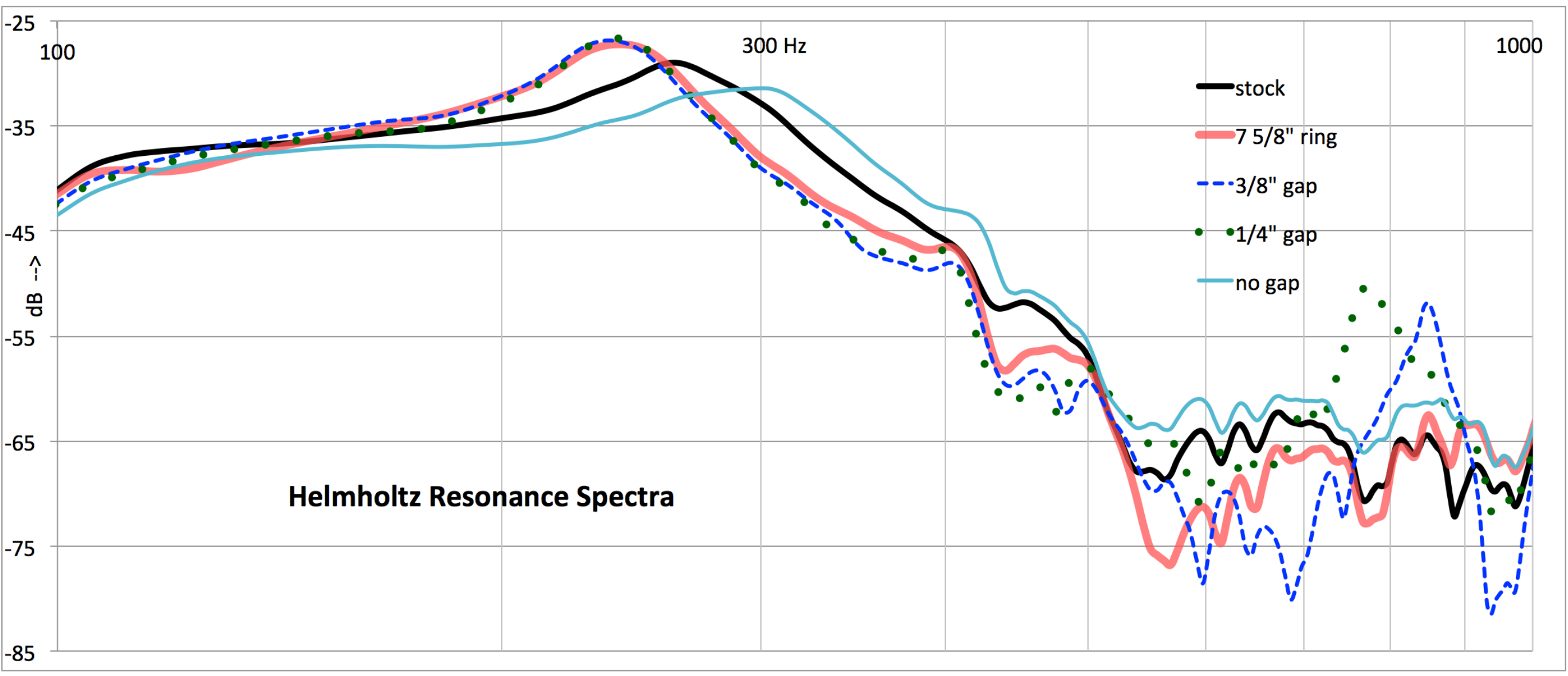}
\caption{Spectra with plywood head with $3''$ speaker and foam belly-back}
\end{figure}

The relations between the stock, $7 {5\over8}''$ and no-gap curves are standard Helmholtz resonator physics.  (As before, referring to the ideal Helmholtz bottle, $A$ is the main volume/neck interface area, and $V_{\text{neck}}$ is the volume of the neck;  let $V$ be the main volume.  Then the ideal Helmholtz frequency is $f_{\text{H}} = {v_s \over {2\pi}} {A \over \sqrt{V V_{\text{neck}}}}$.)  The stock and $7 {5\over8}''$ ring have the same $V$, but the ring has a smaller $A$ and a larger $V_{\text{neck}}$.  The ring and no gap have the same $A$ and $V_{\text{neck}}$, but the no gap has a smaller $V$.  Stock and no gap differ in all three parameters: the no gap has a smaller $V$, a smaller $A$, and a larger $V_{\text{neck}}$.  So the sign of the difference depends on details of the actual values.  However, stock and no gap have approximately equal values of $A/\sqrt{V}$, which accounts for the sign of the observed difference in frequencies.

A very important lesson from these measurements, which was not altogether obvious beforehand, is that the Helmholtz resonances of the $7 {5\over8}''$ ring, ${3\over8}''$ gap, and ${1 \over 4}''$ gap curves, i.e., all of the pots with the same size partial back, are essentially indistinguishable.  That means that the height of the internal cylindrical wall, going from zero up to the height in the standard, finished banjo (i.e., reaching to ${3\over8}''$ from the inner surface of the head) and even beyond by another ${1\over8}''$, does not effect the Helmholtz resonance.  They are all the same --- as if there were no inner wall at all.   The simple Helmholtz resonance picture says that the resonant frequency is independent of the shape of the cavity.  So, apparently, the wall is simply an alteration in the shape.  And these three configurations have the same $V$, $V_{\text{neck}}$, and $A$.  However, one might ask whether there could be a wall sufficiently high that it divides the original cavity into two Helmholtz resonators in series --- just as Rayleigh suggested could arise,\cite{rayleigh} at least for some design.  Apparently, in practice, the answer is no, not for the internal resonator geometry.  There are two obstacles.  Friction becomes an important force with yet smaller gaps.  And the volume of the purported neck between the annulus and the central cylinder is too small relative to the interface area.

\section{Cavity Modes}

The internal wall certainly does something, and that is revealed by a study of the higher frequency cavity modes.  Again, the coupling to the head modes is removed by using a solid ${3\over4}''$ plywood head.  Since these air modes are essentially internal to the pot, the sound hole gap can be eliminated --- allowing for cleaner and clearer resonances.  The sound hole was only crucial to the Helmholtz mode.  So I chose to seal the back with solid plywood.  And that required putting a driving speaker and a recording microphone inside the pot.  That head, speaker, and mic assembly is shown in FIG.~5.

Again, the rim is the Goodtime fit with a Bacon tone ring.  Logarithmic frequency sweeps,  with the small, internally mounted speaker and microphone, yielded the spectra shown in FIG.~8.  The horizontal frequency scale is linear.  ``Stock" refers to the standard rim. ``${3\over8}''$ gap" is the standard internal resonator, whose cylindrical wall is $2 {1\over4}''$ high, which brings it to ${3\over8}''$ from the inner surface of the head.  The task at hand is to understand how the standard internal resonator converts the stock spectrum into the one labeled ${3\over8}''$.  The are no Helmholtz resonances in this configuration because there is no in-and-out air motion.  The lowest closed cavity resonances are the ones shown.

The internal diameter of the pot was $9.84''$.  In terms of the obvious cylindrical coordinates $r$, $\theta$, and $z$, the standard $r$-$\theta$ resonance frequency values are indicated by the solid vertical lines at the bottom of FIG.~8.  (That calculation won't be perfect because it ignores the presence of the speaker, microphone, tone ring, and other hardware inside.)  The dotted lines, first appearing around 2300 Hz, are the calculated frequencies of the additional modes that involve wave components in the $z$ direction for an internal height of $2.97''$.

The key to understanding what is going on is to consider a wall that leaves no gap between itself and the head.  In that case the smaller inner cylinder and the outer annular region are distinct.  A small acoustical coupling between the two was introduced in the form of a ${1\over2}'' \times {1\over2}''$ hole in the internal wall.  The corresponding spectrum is the red ``no gap" curve in FIG.~8.

\begin{figure}[h!]
\includegraphics[width=6.5in]{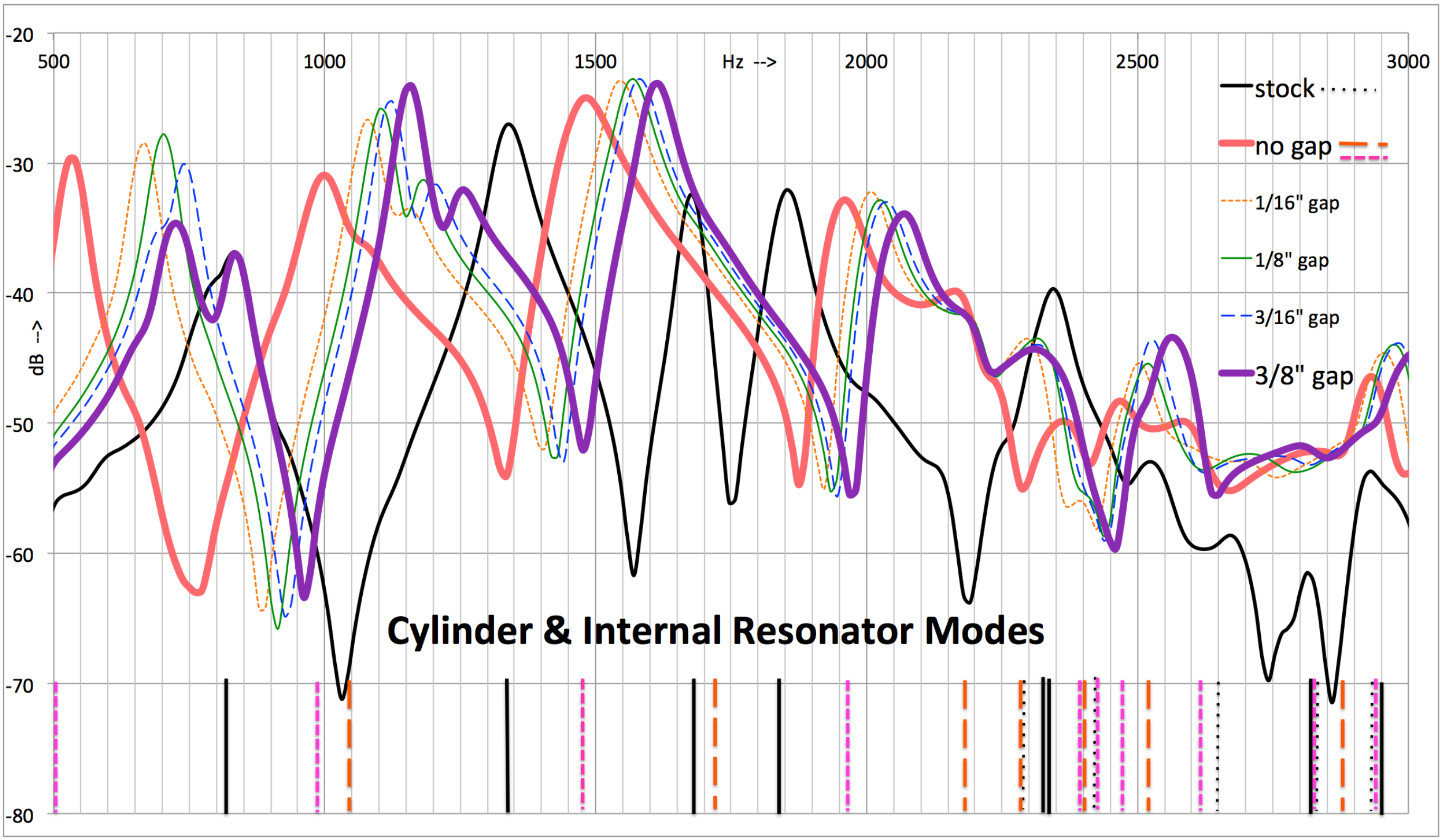}
\caption{Comparison of the stock pot \& the internal resonator with various gaps; lines at the bottom denote frequencies calculated from the actual physical dimensions}
\end{figure}

The calculated resonant frequencies for the no-gap system, under the assumption that the coupling of its two parts is weak enough to ignore, are also indicated by lines at the bottom of FIG.~8.  The long dash lines are the standard cylinder mode frequencies, higher than the solid lines simply by the ratio of the stock diameter to the internal resonator diameter, at least for the $z$-independent, lower frequency range.  The two cylinders have the same height, and the higher frequency contributions from waves in the $z$ direction are added in also.

The calculated mode frequencies of the annular volume are indicated with short dashes  and use the approximation described in section II.  They begin around 500 Hz, which is substantially lower than the lowest mode of the stock cylinder.  Note that the smallest dimension of the annulus is $1.01''$ in the $r$ direction.  That is only first excited around 6700 Hz.

FIG.~8 also displays the measured spectra for intermediate values of the rim-wall-head gap, illustrating how the spectrum evolves continuously from no gap to its final ${3\over8}''$ value.

\section{The Bacon Tone Ring}

Bacon included a metal tone ring in his design.  It sits on the top edge of the wood rim, and the drum head is stretched over it.  A detailed study of its vibrations and their effect on the banjo's sound is presented in ref.~\cite{tone-ring}.  In summary: the ${1\over4}''$ solid diameter core hardens the rim edge to reduce high frequency absorption relative to a pure wood rim.  The two thin flanges stiffen the rim (at the cost of little extra weight) to reduce large rim motions that otherwise absorb low frequencies.  And vibrations of the free horizontal flange absorb some of the vibrational energy that without the flange would have gone into sound.

For the study of air modes, all of the comparisons are made with a Bacon tone ring installed on a single $11''$ Goodtime rim.  Whatever stiffening and vibrating the tone ring does, it does so similarly for the different back and internal geometries.    Only the comparisons in section VI involve an all-wood, standard  Goodtime rim, where it is contrasted with a Bacon-modified Goodtime, i.e., with tone ring and full-size internal resonator installed.  This comparison involves all the effects at once.

\section{Fully assembled banjo plucked and played}

How does a Bacon internal resonator banjo sound and how does it differ from an otherwise identical instrument? People familiar with 5-string banjo music can hear the difference.  However, that does not mean that the differences can be easily discerned from numbers and graphs.  Shown below are spectrographs of individual plucks and frequency analysis of two entire 35 second played samples.  The only obvious differences in FIG.s 9 and 10 are in the enhanced response at the lowest frequencies, as to be expected from the discussion of the air modes of the pots.  There really is no substitute for listening.\cite{bacon-physics}.

The instruments compared are two fully assembled banjos: a totally normal Goodtime and the Bacon-inspired modified Goodtime, i.e., with tone ring and internal $2 {1\over4}''$ high internal resonator.  The strings, heads, and head tensions (as measured by a DrumDial) were the same.  (Deering Goodtime banjos were chosen because they are mass-produced, hand-finished, high-quality instruments that are about as identical as complex constructions made of maple can be.)

FIG.~9 is a spectrograph of four typical single string plucks, with the other stings left free to vibrate.  The microphone was at $20''$ in front of the head.  All plucks were at the second fret.  The first one is the $4^{\text{th}}$ string of the normal Goodtime; the second is the  $4^{\text{th}}$ string of the Bacon-modified Goodtime; the third is the $1^{\text{st}}$ string of the normal Goodtime; and  the fourth is the $1^{\text{st}}$ string of the Bacon-modified Goodtime.  

FIG.~10 presents the frequency spectra for an entire 35 second selection, played and recorded on the two banjos as identically as possible.

\begin{figure}[h!]
\includegraphics[width=6.5in]{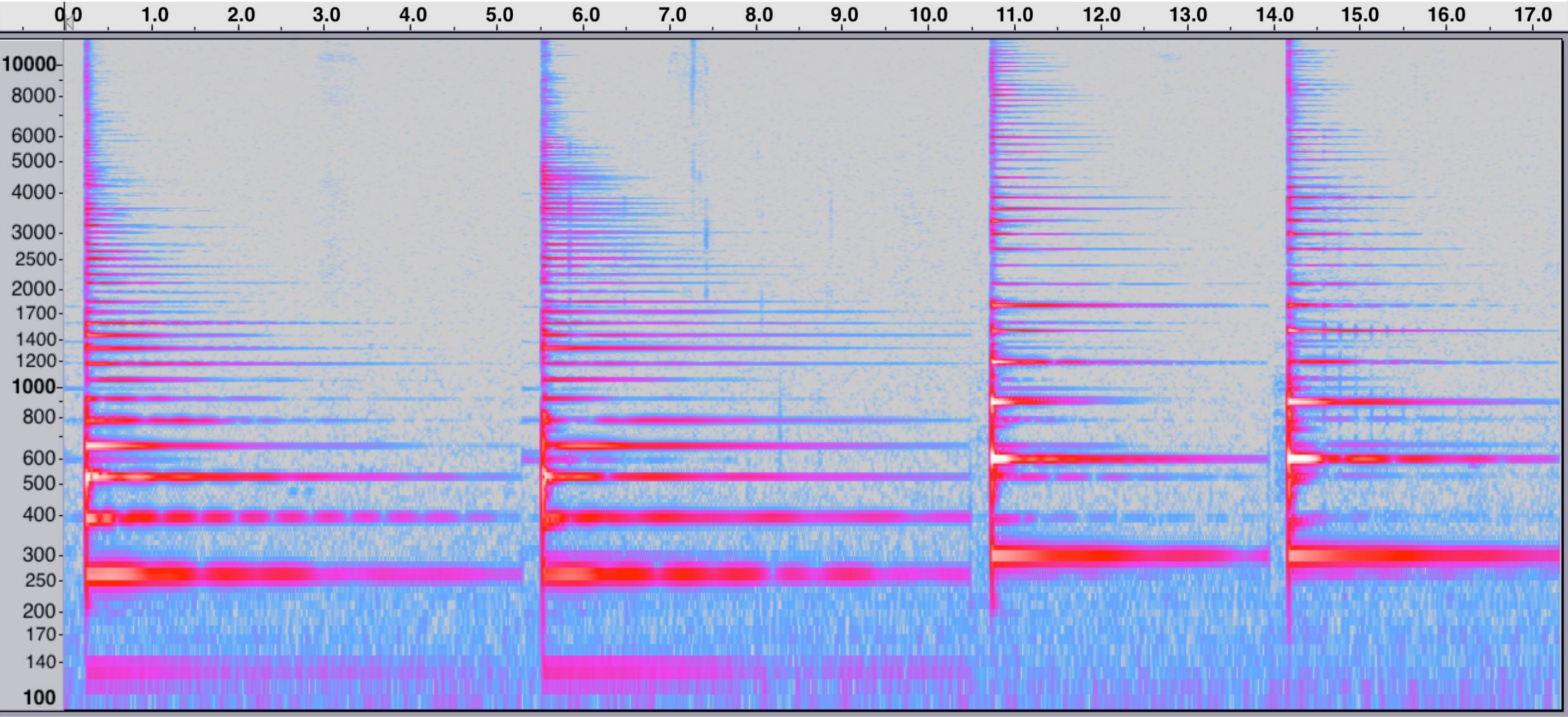}
\caption{Spectrograph of four typical plucks (left to right): $4^{\text{th}}$ string stock, $4^{\text{th}}$ string Bacon-modified, $1^{\text{st}}$ string stock, $1^{\text{st}}$ string Bacon-modified}
\end{figure}

\begin{figure}[h!]
\includegraphics[width=6.5in]{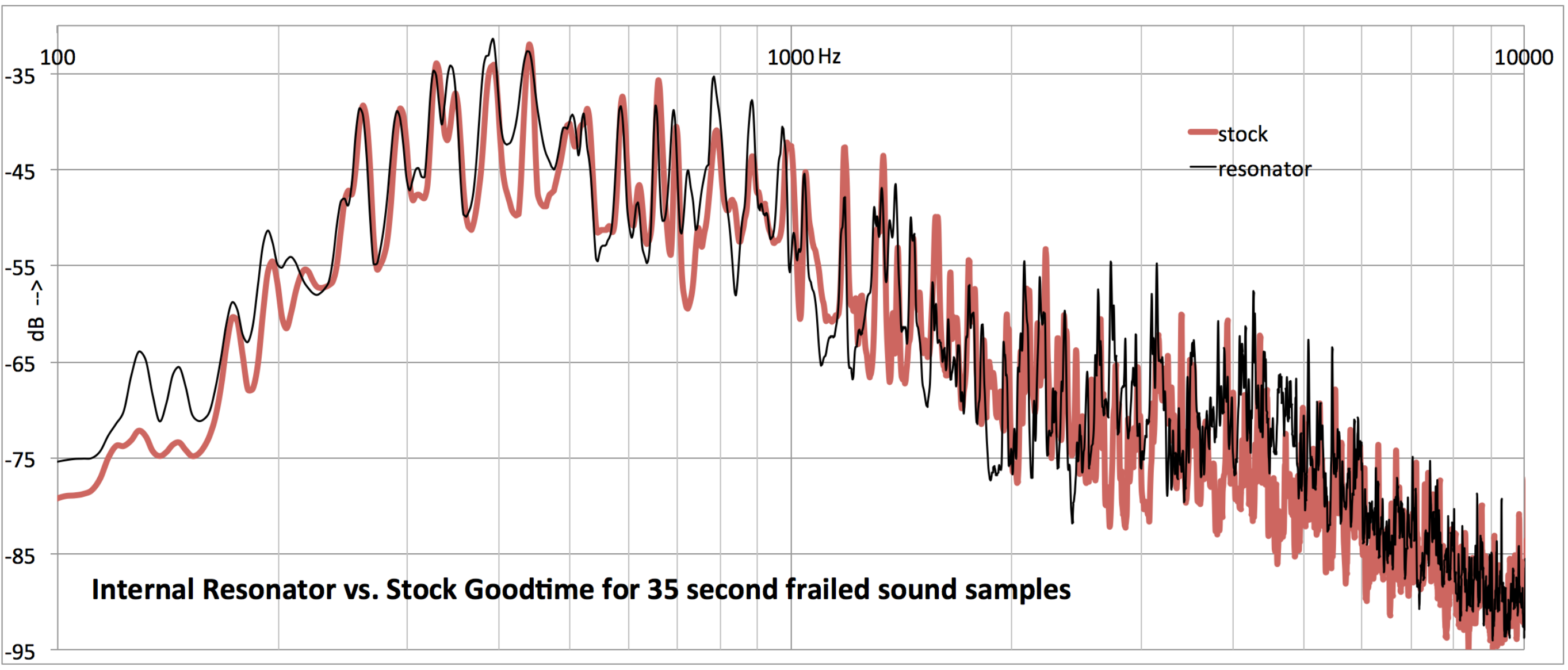}
\caption{Spectra of 35 seconds of frailing on a Goodtime stock {\it vs} internal-resonator-fitted banjo }
\end{figure}

\section{Conclusion}

The cylindrical symmetry typical of banjo design simplifies the analysis of internal air modes.  The ``internal resonator," a modification of the inside geometry of an open-back banjo, defines a smaller central cylinder and an outer annular region.  A simple calculation of air modes of the ideal, separated, two-volume system and a sequence of measurements of increasingly coupled volumes demonstrates that this is a useful picture.  The whole construction produces a richer spectrum, starting at a lower frequency than a simple cylinder of the same outer dimensions.  The lowest frequency corresponds to a Helmholtz resonance that turns out to be insensitive to the division of the internal volume into annular and central cylindrical regions --- at least for practical geometries.  That frequency is lower than the one for a simple open-back banjo of the same outer dimensions because of the internal resonator's impact on the effective sound hole.

On the other hand, the pressure-sensitive drum head/sound board makes intimate contact over its entire area with that internal air.  The head is not particularly close to an ideal membrane nor a plate.  Being very light and flexible, the head has interactions with the air that are neither ignorable nor trivial.  So analysis of its motion presents a serious challenge.   But that motion is what makes almost all of the sound.  A more modest question is a comparison.  How do Bacon's design modifications alter the sound?   A big part of that is clearly the altered dynamics of the air in the pot as studied in this note.  However, that is not the complete story because it is still missing the details of the interaction with the head that drives the internal air motion and also responds to it.

There is a further caveat that goes with any discussion of plucked strings that is based on normal modes and spectral analysis.  The relevant motions are all transients.  Steady state behavior is an approximation --- and not always a good one, especially if one is interested in subtle distinctions.  Sonically interesting features of the transients can be sensitive to small differences in details.\cite{coupled-damped}  This is certainly true of banjos.

\end{document}